\pdfoutput=1 % Required by arxiv
\documentclass[]{birkjour}
\usepackage{physics}
\usepackage{amsthm}
\usepackage{amssymb}
\usepackage{mathtools}  % Needed for floor
\usepackage{csquotes}
\usepackage{orcidlink}
\usepackage{xcolor}
\usepackage{url}
\usepackage{graphicx}
\usepackage{amsmath}
\usepackage{amsfonts}
\usepackage{physics}
\usepackage{hyperref} 
\usepackage{cleveref}
\usepackage{nicefrac}
\usepackage{doi}
\usepackage{slashed}
\usepackage[english]{babel}

\RequirePackage[all]{xy}

\theoremstyle{definition}

\theoremstyle{remark}

\numberwithin{equation}{section}

\newcommand{\BibTeX}{B\kern-0.1emi\kern-0.017emb\kern-0.15em\TeX}
\newcommand{\XYpic}{$\mathrm{X\kern-0.3em\raisebox{-0.18em}{Y}}$-$\mathrm{pic}\,$}

\newcommand{\cl}{C \kern -0.1em \ell}  %%Clifford algebra

\newcommand{\ed}{\end{document}}
\begin{document}

%-------------------------------------------------------------------------
% editorial commands: to be inserted by the editorial office
%
%\firstpage{1} \volume{228} \Copyrightyear{2004} \DOI{003-0001}
%\seriesextraline{This is the Concrete Title of this Book\br H.E. R and S.T.C. W, Eds.}
%\firstpage{1}

\title[The Wigner Little Group for Photons Is a Projective Subalgebra]
 {The Wigner Little Group for Photons Is a Projective Subalgebra}
%----------Author 1
\author[Moab Croft]{Moab Croft \orcidlink{orcid.org/0000-0003-4082-0428}}
\address{%
%Viaduktstr. 42\\
United States of America}
\email{moabphysics@gmail.com}
\author[Hamish Todd]{Hamish Todd}
\address{%
%Viaduktstr. 42\\
United Kingdom}
\email{hamish.todd1@gmail.com}
\author[Edward Corbett]{Edward Corbett}
\address{%
%Viaduktstr. 42\\
United Kingdom}
\email{ga@craft-e.com}

%

%\subjclass{Primary 15A66, 16W50, 20C05, 20C40, 20D15; Secondary 68W30}
%
\sloppy
\date{\today}
%\dedicatory{Last Revised:\\ \today}
%%% ----------------------------------------------------------------------
\begin{abstract}
This paper presents the Geometric Algebra approach to the Wigner little group for photons using the Spacetime Algebra, incorporating a mirror-based view for physical interpretation. The shift from a \textit{point-based view} to a \textit{mirror-based view} is a modern movement that allows for a more intuitive representation of geometric and physical entities, with vectors and their higher-grade counterparts viewed as hyperplanes. This reinterpretation simplifies the implementation of homogeneous representations of geometric objects within the Spacetime Algebra and enables a \textit{relative view} via projective geometry. Then, after utilizing the intrinsic properties of Geometric Algebra, the Wigner little group is seen to induce a projective geometric algebra as a subalgebra of the Spacetime Algebra. However, the dimension-agnostic nature of Geometric Algebra enables the generalization of induced subalgebras to $(1+n)$-dimensional Minkowski geometric algebras, termed \textit{little photon algebras}. The lightlike  transformations (translations) in these little photon algebras are seen to leave invariant the (pseudo)\textit{canonical electromagetic field bivector}. Geometrically, this corresponds to Lorentz transformations that do not change the intersection of the spacelike polarization hyperplane with the lightlike wavevector hyperplane while simultaneously not affecting the lightlike wavevector hyperplane. This provides for a framework that unifies the analysis of symmetries and substructures of point-based Geometric Algebra with mirror-based Geometric Algebra.  
\end{abstract}
\label{Abstract}
%%% ----------------------------------------------------------------------
\maketitle
%%% ----------------------------------------------------------------------
%\tableofcontents
\newpage

\section{Introduction and Motivation}\label{sec0}

The Wigner little group for photons has been studied extensively \cite{Wigner1939,Wigner1987,Han1981,Han1982,Kim2016}. It is known to be isomorphic to the $2$-dimensional special Euclidean group, $\mathrm{SE}(2)$, yet the exact geometric interpretation is lacking. This little group has been related to the cylindrical group through mathematical reasoning, however this applies to understanding the \textit{abstract behavior} of the little group, not the physical result. As such, the Wigner little group for photons will be explored through the use of \textit{Geometric Algebra}.
\begin{figure}
    \centering
    \includegraphics[width=0.4\linewidth]{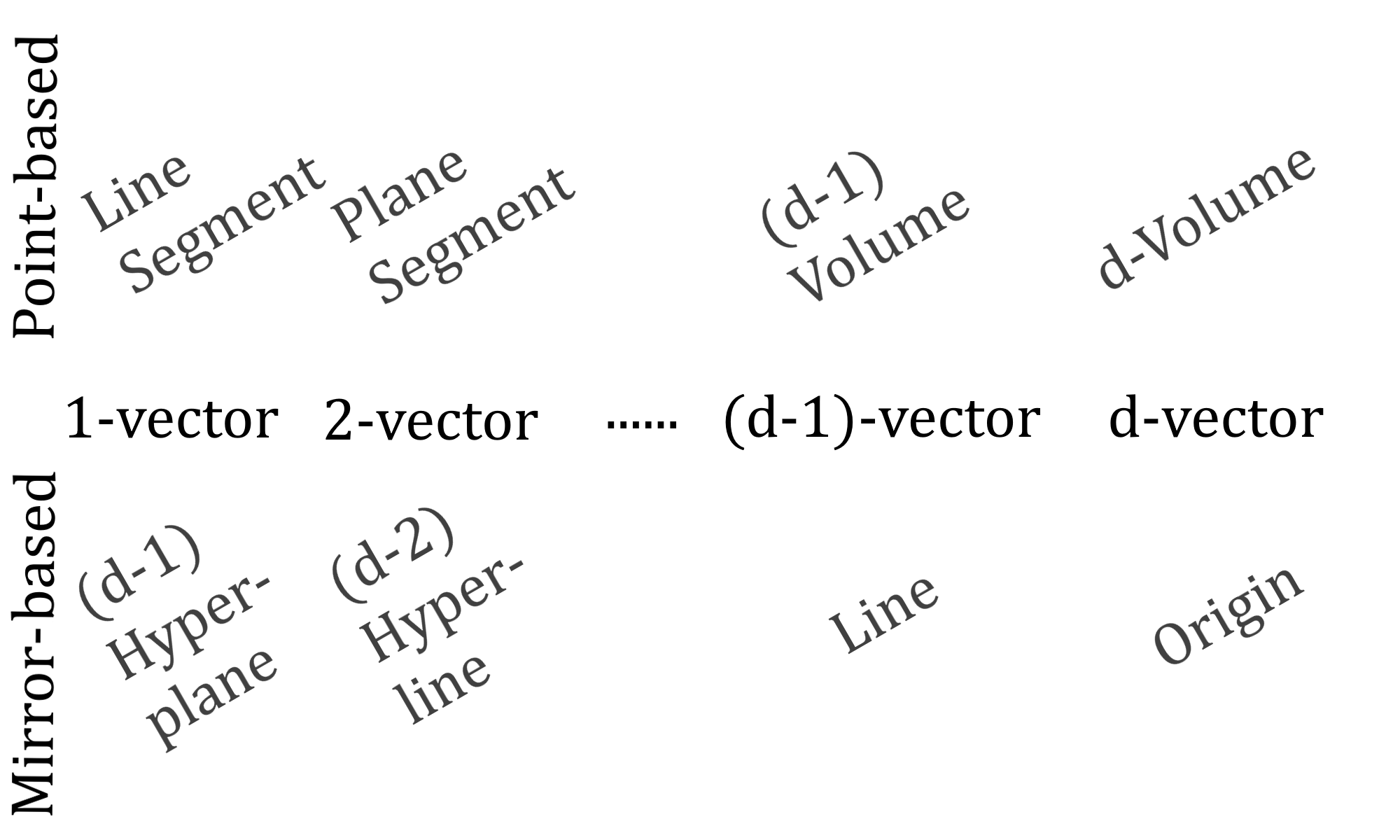}
    \caption{List of geometric interpretation assigned to different grades in both the \textit{point-based view} and \textit{mirror-based view}.}
    \label{FIG:Views}
\end{figure}

The Geometric Algebra approach within physics took its start in 1966 with Hestenes, \cite{Hestenes1966}. In his work, Hestenes employed an interpretation now called the \textit{point-based view} of geometry, where vectors represent directed line segments, bivectors represent oriented plane segments, and so forth. This paper will make use of the \textit{mirror-based view}\footnote{A synonomous term is the \textit{plane-based view} of geometry. However, this is typically used in the context of degenerate geometric (Clifford) algebras.} of geometry, where for a $d$-dimensional geometric algebra modeling $d$-dimensional (pseudo)Euclidean space, vectors represent $(d-1)$-dimensional hyperplanes, bivectors represent $(d-2)$-dimensional hyperlines, and so forth. A list of both interpretations is given in FIG.~\ref{FIG:Views}. The best algebra for describing a single observer in its restframe is the \textit{Spacetime Algebra} \cite{Hestenes1966,Dressel2015,Doran2003,Sobczyk2019}, and it traditionally uses the point-based view. This paper holds the opinion that this view can obfuscate certain geometric truths and also makes for less intuitive realizations of homogeneous representations \cite{Gunn2016,Sokolov2016,Roelfs2023}. In order to fulfill the intuitive capabilities of the mathematics, this paper views it as natural to continue with the mirror-based view of the Spacetime Algebra, which will be elucidated in SEC.~\ref{sec1.2}.

This paper will be structured as follows: The mathematical essentials for the Spacetime Algebra and the Wigner little group for photons will be presented in SEC.~\ref{sec1}. Thereafter it will be shown in SEC.~\ref{sec2} that the little group induces a subalgebra of the Spacetime Algebra that is a projective geometric algebra. Due to the dimension-agnostic nature implicit to Geometric Algebra, it will follow in SEC.~\ref{sec3} that, in Minkowski spaces, Wigner little groups for photons induce subalgebras that are projective geometric algebras. Furthermore, the lightlike Lorentz transformations in these projective subalgebras will be seen to leave the (pseudo)\textit{canonical electromagnetic field bivector} invariant.

\section{Essentials}\label{sec1}

The Spacetime Algebra is traditionally given in the point-based view which does not intuit the existence of geometric objects like hyperplanes, hyperlines, lines, or points as being represented by graded algebraic objects. This section will first present the algebraic essentials of the Spacetime Algebra, and then will present the mirror-based view. Finally, the last segment of this section will present a brief background on Wigner little groups, specifically the Wigner little group for photons, using the traditional Lie and matrix algebra approach. This paper recommends \cite{Dressel2015} for readers who wish for more in-depth materials on the Spacetime Algebra. Likewise, \cite{Roelfs2023} is recommended for readers who wish to learn more about the mirror-based view of geometry. For readers who wish to learn more about the history of the Wigner little group for photons and its traditional formulation, this paper recommends \cite{Kim2016}.

\subsection{The Spacetime Algebra}\label{sec1.2}

The Spacetime Algebra is the real $(1+3)$-dimensional geometric (Clifford) algebra, $\mathbb{G}_{1,3}$, generated by one \textit{unipotent} and three \textit{anti-unipotent} basis vectors,
\begin{equation}\label{EQ:sec1.2: Basis Vectors}
    \gamma_0,\quad \gamma_1, \quad \gamma_2, \quad \gamma_3,
\end{equation}
whose inner products satisfy the (mostly-minus) Minkowski metric, $\gamma_{\mu}\cdot\gamma_{\nu}=\eta_{\mu\nu}$. The basis vectors anticommute with each other and through their outer products generate the basis bivectors,
\begin{equation}\label{EQ:sec1.2: Bivectors}
    \gamma_\mu\wedge \gamma_\nu=(1-\delta_{\mu\nu})\gamma_{\mu\nu}.
\end{equation}
The basis vectors also generate the basis trivectors,
\begin{equation}\label{EQ:sec1.2: Basis Trivectors}
    \gamma_{123}, \quad \gamma_{023}, \quad \gamma_{031}, \quad \gamma_{012}.
\end{equation}
When all four basis vectors are multiplied, the \textit{pseudoscalar} $I=\gamma_{0123}$ is created. The pseudoscalar commutes with all even elements of the algebra and anticommutes with all odd elements. It induces a duality allowing the trivectors to be rewritten as $\gamma_\mu I =-I\gamma_\mu$ and the spacelike (anti-unipotent) bivectors to be rewritten as $\gamma_{j0}I=I\gamma_{j0}$. This is a demonstration of the \textit{complex-like} structure in the Spacetime Algebra. The term complex-\textit{like} is used instead of \textit{complex} because the pseudoscalar does not commute with all elements of the algebra. However, in the subalgebra of even grade elements, $\mathbb{G}_{1,3}^+$, the pseudoscalar does commute with all elements of the algebra and indeed generates a complex structure. Consequently, within this even subalgebra the unit pseudoscalar may be simultaneously treated as the unit imaginary. This fact is also seen in the $2\times2$ complex Pauli matrix representation of the even subalgebra.
\begin{figure}
    \centering
    \includegraphics[width=0.4\linewidth]{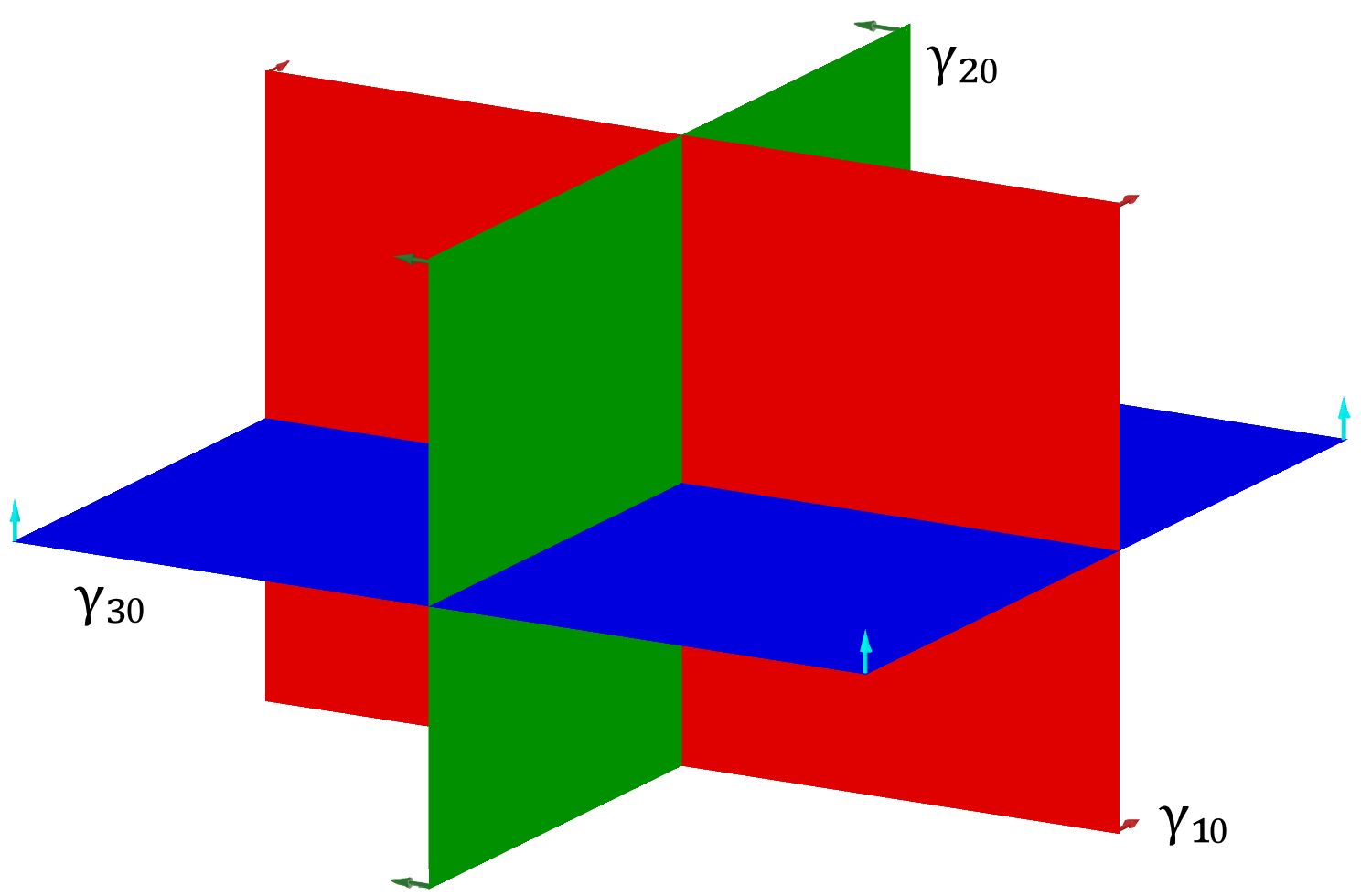}
    \caption{An example of the mirror-based view in the $3$-dimensional relative space $\mathbb{G}_3\approx\mathbb{G}_{1,3}^+$. The red plane is $\gamma_{10}$, the green plane is $\gamma_{20}$, and the blue plane is $\gamma_{30}$. The arrows at the ends of the planes denote the extrinsic orientation of the (hyper)planes.}
    \label{FIG:Basis}
\end{figure}

Geometrically, a vector is a $3$-dimensional hyperplane, and in this paper will be referred to as a \textit{worldvolume}. These have \textit{extrinsic orientation} given by the direction of the vectors in their point-based view. The hyperplane is orthogonal to its extrinsic orientation and passes through the \textit{origin} given by $I$, which conceptualizes the intersection of all worldvolumes, $\gamma_0\wedge\gamma_1\wedge\gamma_2\wedge\gamma_3=I$. An example of this concept in the $3$-dimensional relative space, $\mathbb{G}_3\approx\mathbb{G}_{1,3}^+$ is given in FIG.~\ref{FIG:Basis}. Note that the timelike bivectors are vectors in the $3$-dimensional relative space, $\mathbb{G}_3$, and hence are (hyper)planes.

Both a worldvolume $v$ and a scalar multiple $\alpha v$ describe the same worldvolume due to the homogeneous representation of the Spacetime Algebra. The intersection of two worldvolumes, a bivector, gives a $2$-dimensional hyperline, here called a worldplane. All worldplanes have extrinsic orientation given by the orientation of the bivectors in the point-based view. Likewise, trivectors are the intersection of a worldvolume with a worldplane (e.g. $\gamma_0\wedge\gamma_{12}=\gamma_{012}$), and represent lines, here called \textit{worldlines}. Their orientation
is intrinsic, like a spearline, and they pass through the origin. Worldvolumes, worldplanes, and worldlines can be \textit{timelike} (unipotent), \textit{spacelike}\footnote{Thus timelike, spacelike, and lightlike worldlines exist as geometric objects in the Spacetime Algebra. This paper does not take this as evidence for the physical existence of \textit{tachyons}, but notes it for the sake of generality, akin to \cite{Hidden}.} (anti-unipotent), and \textit{lightlike} (nilpotent). The difference between \textit{extrinsic} and \textit{intrinsic} is quite important for the mirror-based view, because extrinsically and intrinsically oriented objects transform differently under mirroring. An extrinsically oriented hyperplane will swap its direction of orientation under self-mirroring, whereas an intrinsically oriented hyperplane will remain invariant.

Hestenes introduced the \textit{spacetime split} of a vector, $v\gamma_0=v\cdot\gamma_0+v\wedge\gamma_0$, or equivalently,
\begin{equation}\label{EQ:sec1.2: spacetimesplit}
    v = (v_0+v_1\gamma_{10}+v_2\gamma_{20}+v_3\gamma_{30})\gamma_0.
\end{equation}
The elements within the parentheses, $v\gamma_0$, will always be elements of the even subalgebra $\mathbb{G}_{1,3}^+$ and thus correspond to the $3$-dimensional relative space $\mathbb{G}_3$. It is said that $\gamma_0$ induces the relative coefficients of $v$. Like with vectors, $\gamma_0$ can induce relative coefficients for a trivector, $T\gamma_0=T\wedge\gamma_0+T\cdot\gamma_0$, or equivalently,
\begin{equation}\label{EQ:sec1.2: spacetimesplit2}
    T = (T_0I^{-1}+T_1\gamma_{23}+T_2\gamma_{31}+T_3\gamma_{12})\gamma_0.
\end{equation}
In the mirror-based view, worldvolumes (vectors) are most appropriate for describing physical quantities related to direction and orientation, while worldlines (trivectors) are most appropriate for describing inertial trajectories through the origin $I$. For example, a worldvolume $p$ describes \textit{relativistic momentum} with coefficients induced by $\gamma_0$ following EQ.~\ref{EQ:sec1.2: spacetimesplit}, while a worldline $X$ describes \textit{relativistic position} with coefficients induced by $\gamma_0$ following EQ.~\ref{EQ:sec1.2: spacetimesplit2}. 

\subsubsection{The Relative View of the Spacetime Algebra}\label{relativeview}

One can break into a relative frame through the isomorphism between relative $3$-dimensional space and the even subalgebra of the Spacetime Algebra, $\mathbb{G}_3\approx\mathbb{G}_{1,3}^+$, by using the spacetime splits of EQ.~\ref{EQ:sec1.2: spacetimesplit} and EQ.~\ref{EQ:sec1.2: spacetimesplit2}. However, some geometric information is necessarily lost in such a process due to the difference in bases between $\mathbb{G}_3$ and $\mathbb{G}_{1,3}$. For example, a lightlike worldvolume $\gamma_0+\gamma_3$ does not correspond to a primitive geometric object like a hyperplane within $\mathbb{G}_3$. The geometric analysis of the Wigner little group for photons will heavily depend upon such lightlike worldvolumes, so an alternative method must be used. This paper will therefore apply the projective methods of \cite{Sokolov2016} to create a \textit{relative view} of the Spacetime Algebra. The contents of \cite{Sokolov2016} cannot be fully discussed within this paper, but the approach will be summarized and its appropriate visualization will be explained.

To more easily explain the approach, this subsection will use the Lightcone Algebra, $\mathbb{G}_{1,2}$. All vectors of the Lightcone Algebra are $2$-dimensional planes, called worldplanes. Thus all bivectors are $1$-dimensional lines, called worldlines. And the basis trivector is the origin. For the sake of simplicity, this subsection will ignore orientation. The basis worldplanes look just like FIG.~\ref{FIG:Basis}, although the red and green planes would be anti-unipotent in the Lightcone Algebra rather than unipotent like in $\mathbb{G}_3$. The projective methods of \cite{Sokolov2016} map $d$-dimensional space to $(d-1)$-dimensional space, so the relative view of the Lightcone Algebra will be $2$-dimensional. To create this relative view (which is the \textit{hyperbolic} projective space in \cite{Sokolov2016}), a plane parallel to the timelike basis worldplane is chosen, and the intersections of this plane with all geometric elements in the Lightcone Algebra form the relative view. This is demonstrated in FIG.~\ref{FIG:sk}. So, worldplanes in the Lightcone Algebra are seen as \textit{lines} in the relative view, and worldlines are seen as \textit{points}. In the relative view, the origin is a point outside the projective plane and therefore is seen to be infinitely\footnote{The lead author of this paper advocates for the phrase, "Nowhere yet everywhere" as a tool for interpreting the origin in the relative view.} far away. Notice, in FIG.~\ref{FIG:sk}, that the lightlike worldvolume $k$ is still a primitive geometric object within the relative view, which will allow for a geometric analysis of the Wigner little group for photons. It will be assumed in this paper that the coefficients of relative view objects are induced by $\gamma_0$ according to EQ.~\ref{EQ:sec1.2: spacetimesplit} and EQ.~\ref{EQ:sec1.2: spacetimesplit2} in order to relate the concept of the spacetime split to the concept of the relative view.

\subsection{The Wigner Little Group for Photons}\label{sec1.3}

A \textit{Wigner little group} for a particle with momentum $p$ is the group of transformations under which $p$ is invariant. For massive particles, this is $\mathrm{SO}(3)$. For massless particles, this is $\mathrm{SE}(2)$, and the group is called the \textit{Wigner little group for photons}. 

The Wigner little group for photons is traditionally presented using matrix algebra, so this section presents the definition in its matrix algebra form. The (orthochronous) Lorentz group $\mathrm{SO}^+(1,3)$ is generated by the set of rotations $\{J_1,J_2,J_3\}$ about the orthogonal axes of $3$-dimensional Euclidean space and the set of boosts $\{K_1,K_2,K_3\}$ in the three orthogonal directions of this space. Specifically, these generators satisfy the Lie algebra commutation relations,
\begin{equation}\label{EQ:sec1.3: Lie Commutation SO(1,3)}
    [J_j,J_k]=i\varepsilon_{jkl}J_{l},\quad\quad [J_j, K_k]=i\varepsilon_{jkl}K_l, \quad\quad [K_j,K_k]=-i\varepsilon_{jkl}J_l.
\end{equation}
The Wigner little group for photons is then found by considering a subgroup formed for a lightlike particle traveling along the $z$-axis,
\begin{equation}\label{EQ:sec1.3: LGP Generators}
    J_3,\quad\quad N_1 = K_1 - J_2, \quad\quad N_2 = K_2 + J_1.
\end{equation}
These generators satisfy a new set of Lie algebra commutation relations,
\begin{equation}\label{EQ:sec1.3: Lie Commutation SE(2)}
    [N_1,N_2]=0,\quad\quad [J_3, N_1]=iN_2, \quad\quad [J_3,N_2]=-iN_1,
\end{equation}
which forms the Lie algebra for the $2$-dimensional special Euclidean group, $\mathrm{SE}(2)$.

\section{Projective Subalgebras}\label{sec2}

The Spacetime Algebra, $\mathbb{G}_{1,3}$, is algebraically equivalent to the conformal algebra of $2$-dimensional anti-Euclidean space, so the existence of a projective subalgebra for $2$-dimensional anti-Euclidean space will be unsurprising \cite{Hrdina2021}. The anti-Euclidean subalgebra, $\mathbb{G}_{0,2,1}$, might sound unrelated to the Euclidean subalgebra, $\mathbb{G}_{2,0,1}$, as the two algebras are not isomorphic. However, their spin groups, respectively $\mathrm{Spin}(0,2,1)$ and $\mathrm{Spin}(2,0,1)$, \textit{are} isomorphic and double cover the $2$-dimensional Euclidean group \cite{Roelfs2023}, $\mathrm{SE}(2)$. Because of this fact, this paper will continue to use the mostly-minus metric for the Spacetime Algebra. 

\subsection{The Wigner Little Group as a Subalgebra}\label{sec2.1}

Within the Spacetime Algebra, the generators and Lie algebra commutation relations of the Lorentz group take the form
\begin{equation}\label{EQ:sec2.1: STA Lie Alg SO(1,3)}
    \gamma_{jk}\times\gamma_{kl}=\varepsilon_{jkl}\gamma_{lj}, \quad\quad \gamma_{jk}\times \gamma_{k0}=-\varepsilon_{jkl}\gamma_{kl}I, \quad\quad \gamma_{j0}\times\gamma_{k0}=-\varepsilon_{jkl}\gamma_{l0}I,
\end{equation}
which generate the double covering spin group, $\mathrm{Spin}(1,3)$, and use the commutator product $a\times b = (ab-ba)/2$. It follows that the generators of the Wigner little group for photons are 
\begin{equation}\label{EQ:sec2.1: LGP Generators}
    J_3 = \gamma_{12}, \quad\quad N_1 = \gamma_{10}-\gamma_{31}, \quad\quad N_2 = \gamma_{20}+\gamma_{23},
\end{equation}
and satisfy the Lie algebra commutation relations
\begin{equation}\label{EQ:sec2.1: LGP Lie}
    N_1\times N_2=0, \quad J_3\times N_1=N_2, \quad J_3\times N_2=-N_1.
\end{equation}
Noting that $N_1=N_2 I$, it is possible to reduce EQ.~\ref{EQ:sec2.1: LGP Lie} to a single commutation relation,
\begin{equation}\label{EQ:sec2.1: LGP Lie Better}
    J_3\times \left(N_2e^{\frac{\pi}{4}I}\right) \propto J_3\times(N_2(1+I))
\end{equation}
\[
=J_3\times N_2 + I(J_3\times N_2),
\]
which highlights the aforementioned complex-like structure of the algebra. Akin to non-Geometric Algebra approaches \cite{Han1981,Han1982,Kim2016}, a lightlike (translational) Lorent rotor can be constructed,
\begin{equation}\label{EQ:sec2.1: Lorentz Rotor}
    \Lambda = e^{-\frac{1}{2}\theta N_2}=1-\frac{1}{2}\theta N_2,
\end{equation}
where $\theta=\alpha+\beta I$ is a complex-like number comprised of a scalar and a multiple of $I$. Transforming a complex-like vector potential\footnote{The trivector potential contribution $bI$, considered for mathematical generality, results from magnetic charges as detailed in \cite{Dressel2015}.} (vector plus trivector) $z=a+bI=\gamma_0(a_0+b_0I)+\gamma_1(a_1+b_1I)+\gamma_2(a_2+b_2I)+\gamma_3(a_3+b_3I)$,
\[
\Lambda z \Lambda^{-1} 
\]\[
= e^{-\frac{1}{2}\theta N_2} [\gamma_0(a_0+b_0I)+\gamma_1(a_1+b_1I)+\gamma_2(a_2+b_2I)+\gamma_3(a_3+b_3I)]e^{\frac{1}{2}\theta N_2}
\]
\begin{equation}\label{EQ:sec2.1: Transforming z}
    = (\gamma_0+\gamma_3)(a_0-\alpha a_2 - \beta a_1+(b_0-\alpha b_2 - \beta b_1)I) + \gamma_1(a_1+b_1I)+\gamma_2(a_2+b_2I),
\end{equation}
gives a generalization of the gauge transformations previously found in \cite{Wigner1939,Wigner1987,Han1981,Han1982,Kim2016} if and only if the gauge condition $(a_0+b_0I)=(a_3+b_3I)$ is satisfied. Should magnetic charge not be considered, the gauge transformation returns to the form in \cite{Wigner1939,Wigner1987,Han1981,Han1982,Kim2016},
\begin{equation}\label{EQ:sec2.1: Transforming z no Magnetics}
    \Lambda a \Lambda^{-1} = (a_0-\alpha a_2 - \beta a_1)(\gamma_0+\gamma_3) + a_1\gamma_1+a_2\gamma_2.
\end{equation}
In vacuum, the form of EQ.~\ref{EQ:sec2.1: Transforming z no Magnetics} is related to the form of EQ.~\ref{EQ:sec2.1: Transforming z} by a gauge transformation. This is a demonstration of \textit{vacuum representation degeneracy}, which has been experimentally and theoretically demonstrated in \cite{Neugebauer2018,Bliokh2014}.

\subsection{Little Photon Algebra}

The generators in EQ.~\ref{EQ:sec1.3: LGP Generators} are themselves generated by the basis worldvolumes
\begin{equation}\label{EQ:sec2.1: LGP Worldvolumes}
    \gamma_0 + \gamma_3,\quad\quad \gamma_1,\quad\quad \gamma_2.
\end{equation}
These serve as the basis for a $2$-dimensional anti-Euclidean projective geometric algebra. Because this algebra is related to the Wigner little group for photons, it is denoted $\mathbb{W}_{1,3}(\gamma_0+\gamma_3)$ which is isomorphic to $\mathbb{G}_{0,2,1}$ and called a \textit{little photon algebra} of $(1+3)$-dimensional spacetime. As said before, the spin group of this little photon algebra, $\mathrm{Spin(0,2,1)}$ is a double cover of the $2$-dimensional special Euclidean group, $\mathrm{SE}(2)$. This algebra has one nilpotent basis vector, $e_0=\gamma_0+\gamma_3$, and two spacelike basis vectors, $e_1=\gamma_1$ and $e_2=\gamma_2$, satisfying
\begin{equation}\label{EQ:sec2.1: PGA Inner}
    e_\mu\cdot e_\nu =\begin{cases}
        0 & \leftrightarrow \quad  \mu=\nu=0\\
        -\delta_{\mu\nu} & \leftrightarrow \quad  \mu=\nu\neq0
    \end{cases}.
\end{equation}
Of course, EQ.~\ref{EQ:sec2.1: PGA Inner} can be satisfied by \textit{any} lightlike worldvolume and its two corresponding mutually orthogonal spacelike worldvolumes. Thus it can be said that a lightlike vector $k$ \textit{induces} a projective geometric algebra (which is the little photon algebra), $\mathbb{W}_{1,3}(k)$.

\section{Discussion}\label{sec3}

\subsection{Physical Interpretation of $\mathbb{W}_{1,2}$}\label{sec3.1}

To motivate the physical interpretation of $\mathbb{W}_{1,3}$, it is easiest to lower the spatial dimension by one and look at $\mathbb{W}_{1,2}\approx\mathbb{G}_{0,1,1}$, the subalgebra of the Lightcone Algebra $\mathbb{G}_{1,2}$. Its basis vectors, $\mu_0$, $\mu_1$, and $\mu_2$ satisfy the corresponding (mostly-minus) Minkowski metric, and represent extrinsically oriented \textit{worldplanes} passing through the origin. Continuing, bivectors represent extrinsically oriented worldlines through the origin, and the pseudoscalar $i=\mu_{012}$ is the origin. This section, and proceeding sections, will rely on the \textit{relative view} described in SEC.~\ref{relativeview}. Note that the basis for $\mathbb{W}_{1,2}(\mu_0+\mu_2)$ is $e_0=\mu_0+\mu_2$ and $e_1=\mu_1$.
\begin{figure}
    \centering
    \includegraphics[width=0.5\linewidth]{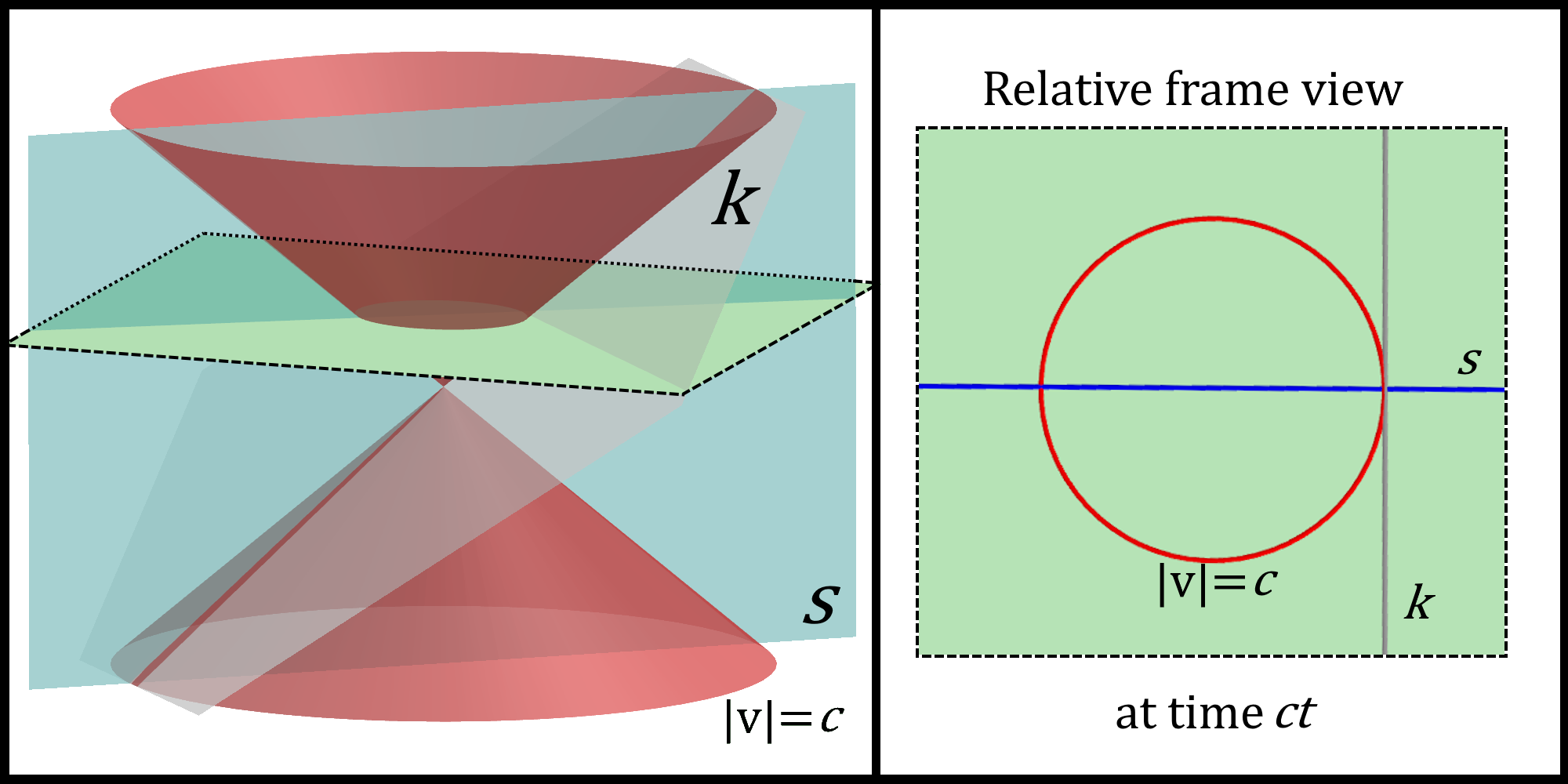}
    \caption{Visualization in $\mathbb{G}_{1,2}$. Left: The lightlike worldplane (gray), $k$, lies tangent to the lightcone (red), $|v|=c$, with the spacelike worldplane (blue), $s$. Right: An observer's relative frame at time $ct$, where the lightlike worldplane (gray), $k$, looks like a line tangent to the light-circle (red), $|v|=c$, with the spacelike worldplane (blue), $s$, which looks like a line orthogonal to $k$ intersecting on the light-circle.}
    \label{FIG:sk}
\end{figure}

Consider the lightlike worldplane $k$ along with the orthogonal spacelike worldplane $s$ satisfying
\begin{equation}\label{EQ:sec3.1: Orthogonal s}
    s\cdot k = 0.
\end{equation}
Without considering the relative view, the lightlike worldplane $k$ is tangent to the lightcone, and in the relative view, it is a line tangent to the light-circle. Similarly, the spacelike worldplane $s$ is orthogonal to $k$ and intersects it along the lightcone without considering the relative view, while in the relative view, it is a line orthogonal to the line $k$ and intersecting it on the light-circle. This is demonstrated in FIG.~\ref{FIG:sk}. After considering the $(1+2)$-dimensional equivalent to EQ.~\ref{EQ:sec2.1: Lorentz Rotor},
\begin{equation}\label{EQ:sec3.1: 1+2 Rotor}
    \Lambda = e^{-\frac{1}{2}\theta e_{10}}=1-\frac{1}{2}\theta e_{10},
\end{equation}
(with real number $\theta$) applying it to the worldplanes $s=s_1e_1$ and $k=k_0e_0$,
\begin{equation}\label{EQ:sec3.1: Covariant Transformation of s and k}
    \Lambda s \Lambda^{-1}=s-\theta s_1e_0, \quad\quad \Lambda k \Lambda^{-1}=k,
\end{equation}
and then considering the intersection of these transformed worldplanes,
\begin{equation}\label{EQ:sec3.1: Intersection}
    (s-\theta s_1e_0)\wedge k = sk,
\end{equation}
it is found that the worldline $sk$ is invariant under the lightlike Lorentz transformation of EQ.~\ref{EQ:sec3.1: 1+2 Rotor}. Without considering the relative view, these transformations then geometrically correspond to the transformation of \textit{pure} spacelike $s$ into \textit{impure} spacelike $\Lambda s \Lambda^{-1}$. In the relative view, this corresponds to the transformation of $s$ which passes through the relative view's (projective) origin $\mu_{12}$ to $\Lambda s \Lambda^{-1}$ which does \textit{not} pass through the relative view's origin $\mu_{12}$. Therefore, \textit{any} transformation in the little photon algebra $\mathbb{W}_{1,2}(k)$ leaves the worldline $sk$ invariant.

\subsection{Physical Interpretation of $\mathbb{W}_{1,3}$}\label{sec3.2}

\begin{figure}
    \centering
\includegraphics[width=0.5\linewidth]{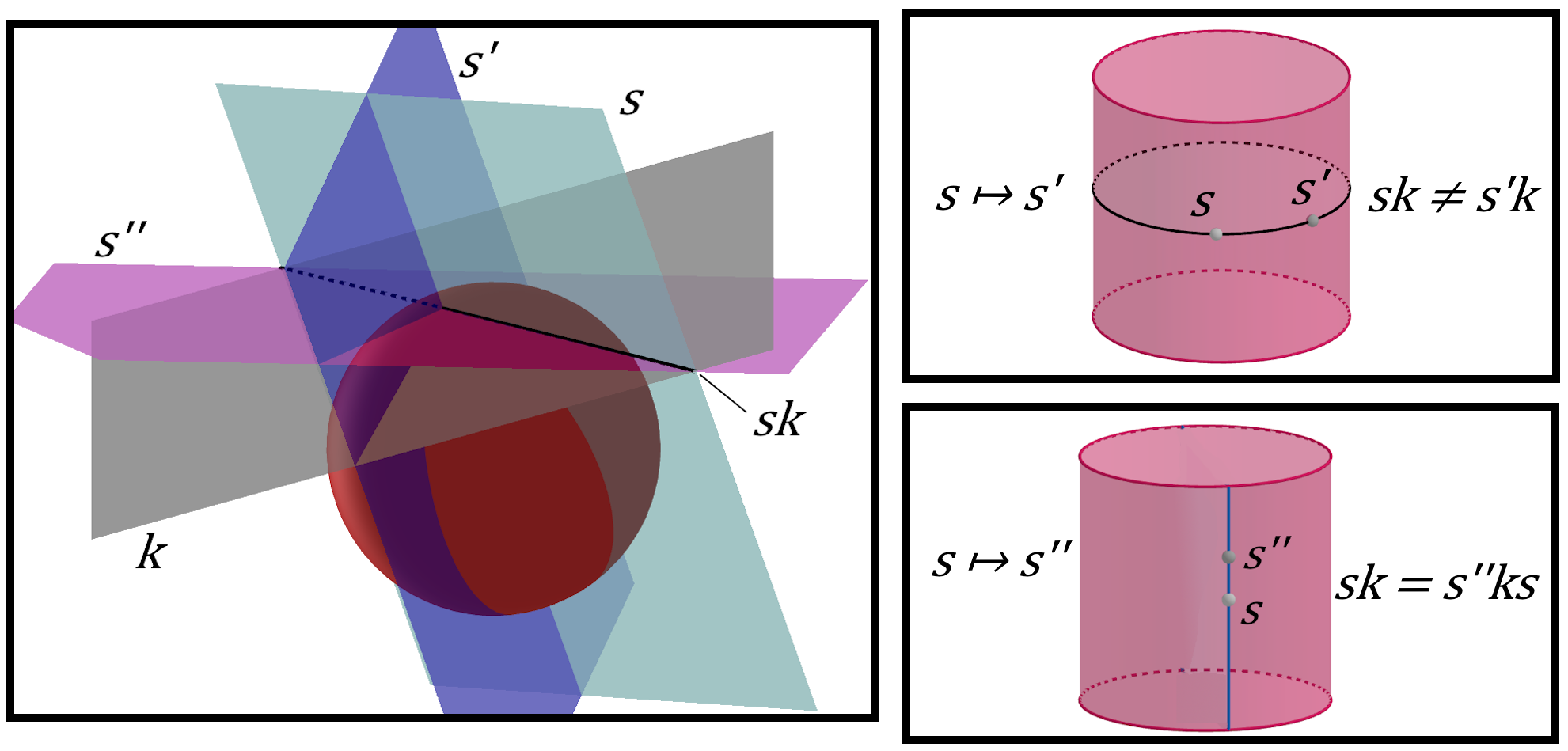}
    \caption{Left: Visualization in the relative view of the Spacetime Algebra. Spacelike $s$ (light blue) is rotated about $\gamma_{12}=e_{12}$ into $s'$ (dark blue), and separately Lorentz transformed into $s''$ (pink) by the rotor in EQ.~\ref{EQ:sec2.1: Lorentz Rotor}. The canonical bivector $sk$ (black line) is unchanged by $s\mapsto s''$. The red sphere represents the light-sphere at moment $ct$. Top right: In the cylindrical group, $s\mapsto s'$ is equivalent to moving on a circle orthogonal to the height, and does not leave $sk$ invariant. Bottom right: In the cylindrical group, $s\mapsto s''$ is equivalent to moving vertically, and leaves $sk$ invariant.}
    \label{FIG:Cylinder}
\end{figure}

Returning to $(1+3)$-dimensional spacetime, consider a lightlike worldvolume $k=k_0(\gamma_0+\gamma_3)$ that induces $\mathbb{W}_{1,3}(k)$. Then consider some $s$ that satisfies EQ.~\ref{EQ:sec3.1: Orthogonal s}. The lightlike $k$ and the spacelike $s$ can respectively be associated with the electromagnetic wavevector and the polarization vector in \cite{Dressel2015}. Thus $sk$ is the (pseudo)\textit{canonical electromagnetic field bivector}. It is called (pseudo)canonical because there are many $s$ that satisfy EQ.~\ref{EQ:sec3.1: Orthogonal s} in $(1+3)$-dimensional spacetime. For example, if $s=s_1\gamma_1+s_2\gamma_2$, then $s$ is \textit{not} invariant under the rotation $s'=e^{-\frac{1}{2}\alpha\gamma_{12}}se^{\frac{1}{2}\alpha\gamma_{12}}$, but the rotated $s'$ remains orthogonal to $k$. It follows that both $s$ and $k$ are elements of the little photon algebra $\mathbb{W}_{1,3}(k)$, and that the rotation generated by $e_{12}=\gamma_{12}$ does \textit{not} leave $sk$ invariant, and instead rotates into $s'k$. However, the Lorentz transformation in EQ.~\ref{EQ:sec2.1: Lorentz Rotor} does leave $sk$ invariant. First, $s$ and $k$ are transformed,
\begin{equation}\label{EQ:sec3.2: Covariant Transformation of s and k}
    \Lambda s \Lambda^{-1}=s-(\alpha s_2 + \beta s_1)e_0, \quad\quad \Lambda k \Lambda^{-1}=k.
\end{equation}
Then, the intersection is considered,
\begin{equation}\label{EQ:sec3.2: Intersection}
    (s-(\alpha s_2 + \beta s_1)e_0)\wedge k = sk,
\end{equation}
proving that the worldplane $sk$ is invariant under lightlike Lorentz transformations. The fact that the canonical bivector $sk$ \textit{does} change under rotations generated by $e_{12}$ is equivalent to moving a point in a circle (orthogonal to the height of a cylinder) within the \textit{cylindrical group}. Furthermore, the fact that lightlike transformations \textit{do not} change $sk$ is equivalent to moving a point vertically within the cylindrical group. This is what was found in \cite{Wigner1987,Kim2016}. This idea is demonstrated in FIG.~\ref{FIG:Cylinder} within the relative view of the Spacetime Algebra. Thus, there is finally a physical correspondence to moving within the abstract cylindrical group: Horizontal translations in the cylindrical group rotate the \textit{polarization worldvolume} $s$ about $e_{12}$ in the Spacetime Algebra and in $\mathbb{W}_{1,3}(k)$; Vertical translations in the cylindrical group will preserve the (pseudo)\textit{canonical electromagnetic field bivector} $sk$ in the Spacetime Algera and \textit{translate} in $\mathbb{W}_{1,3}(k)$. All such transformations leave the \textit{lightlike worldvolume} $k$ invariant by nature of the Wigner little group for photons.

\subsection{A Note on $(1+n)$-dimensional Minkowski Space}\label{sec3.3}

It was shown in SEC.~\ref{sec2} that the Wigner little group for photons induced a projective algebra which functions as a subalgebra of a Minkowski geometric algebra. Therefore, the Wigner little group for photons in $(1+n)$-dimensional Minkowski space will have analogous results to what has already been shown. This is a fact similarly demonstrated by the relationship between projective and conformal geometric algebras in \cite{Hrdina2021}. It was mentioned that the choice of metric is unimportant, so it follows that for the Minkowski geometric algebras, $\mathbb{G}_{1,n}$ or $\mathbb{G}_{n,1}$, the \textit{little photon algebras} are
\begin{equation}\label{EQ:DIS:n: 1+n & n+1}
    \mathbb{W}_{1,n}\approx\mathbb{G}_{0,n-1,1}\subset\mathbb{G}_{1,n},\quad\quad\mathbb{W}_{n,1}\approx\mathbb{G}_{n-1,0,1}\subset\mathbb{G}_{n,1}.
\end{equation}
Indeed, this paper advocates for the term little photon algebra to emphasize the connection between the projective geometric subalgebra and the Wigner little group for photons. it is very important to reiterate that there are infinitely many little algebras, because there are infinitely many lightlike vectors. This is similarly stated in \cite{Hrdina2021,sobszyk2024}. To capture this idea, it is perhaps more reasonable to introduce a little algebra dependent upon $k$, $\mathbb{W}(k)$. Thus it is more appropriate to say that for every lightlike $k$ in $\mathbb{G}_{1,n}$ or $\mathbb{G}_{n,1}$, then respectively there are corresponding little algebras,
\begin{equation}\label{EQ:DIS:n: 1+n & n+1}
    \mathbb{W}_{1,n}(k)\approx\mathbb{G}_{0,n-1,1}\subset\mathbb{G}_{1,n},\quad\quad\mathbb{W}_{n,1}(k)\approx\mathbb{G}_{n-1,0,1}\subset\mathbb{G}_{n,1},
\end{equation}
whose lightlike transformations leave invariant the bivector $sk$ satisfying EQ.~\ref{EQ:sec3.1: Orthogonal s}.

\subsubsection{Little Photon Algebras Live on the Lightcone}

It is worth emphasizing that for a given Minkowski geometric algebra, $\mathbb{G}_{1,n}$, the little photon algebra $\mathbb{W}_{1,n}(k)$ lives on the lightcone. That is, the lightlike (nilpotent) basis vector of the little photon algebra is always tangent to the lightcone, and the spacelike (anti-unipotent) bases are orthogonal to the lightlike basis vector but intersect it along the lightcone. This is clearly seen in both FIG.~\ref{FIG:sk} and FIG.~\ref{FIG:Cylinder}.

\subsection{Conclusion}\label{sec4}

This paper covered the Geometric Algebra approach to the Wigner little group for photons in the Spacetime Algebra, and used the mirror-based view for physical interpretation. By leveraging the intrinsic properties of Geometric Algebra, the Wigner little group was demonstrated to induce a projective geometric algebra as a subalgebra of the Spacetime Algebra. The paper first revisited the history of the Geometric Algebra approach and emphasized the shift from a \textit{point-based view} of geometry to a \textit{mirror-based view}. This shift allowed for a more intuitive understanding and representation of geometric and physical entities within the algebra. The mirror-based view of the Spacetime Algebra gives vectors and their higher grade counterparts as hyperplanes, providing an intuitive framework for representing physical phenomena such as worldlines in a geometrically intuitive manner. This reinterpretation of the Spacetime Algebra facilitated an easier implementation of the homogeneous representation of geometric entities (which enables the relative view), and the insights gained from this approach gave a one-to-one correspondence between the transformations' interpretations in the cylindrical group, the Spacetime Algebra, and the appropriate little photon algebra. Due to the dimension-agnostic nature of the Geometric Algebra approach, similarly induced subalgebras were generalized to $(1+n)$-dimensional Minkowski geometric algebras (analogous to \cite{Hrdina2021}), and such generalized induced subalgebras were coined as \textit{little photon algebras}. Moreover, the lightlike transformations of these little algebras, $\mathbb{W}(k)$, leave the (pseudo)\textit{canonical electromagnetic field bivector} $sk$ invariant. This suggests a
powerful unifying framework for analyzing symmetries and substructures within Geometric Algebra-based physics.

% ------------------------------------------------------------------------

\section*{Acknowledgements}
The authors would like to thank friends and family for their support during this research. They would also like to extend a special thanks to the editor and reviewers for their invaluable advice and time.

\section*{Original Research}
This paper is a pre-final-acceptance-proof version of a publication in \textbf{Springer Link} \textit{Advances in Applied Clifford Algebras}. The official DOI is \textit{10.1007/s00006-025-01369-8}. Please cite the official release.

\bibliographystyle{unsrt}
\bibliography{myBibLib}

\begin{thebibliography}{10}

\bibitem{Wigner1939}
E.~Wigner.
\newblock On unitary representations of the inhomogeneous lorentz group.
\newblock {\em Ann. Math.}, 40:149--204, 1939.

\bibitem{Wigner1987}
Y.~S. Kim and E.~Wigner.
\newblock Cylindrical group and massless particles.
\newblock {\em Journal of Mathematical Physics}, 28, 1987.

\bibitem{Han1981}
D.~Han and Y.S. Kim.
\newblock Little group for photons and gauge transformations.
\newblock {\em Am. J. Phys.}, 49:348--351, 1981.

\bibitem{Han1982}
Y.~Kim D.~Han and D.~Son.
\newblock E(2)-like little group for massless particles and neutrino polarization as a consequence of gauge invariance.
\newblock {\em Physical Review D}, 1982.

\bibitem{Kim2016}
Y.S. Kim.
\newblock Symmetries of massive and massless neutrinos, 2016.

\bibitem{Hestenes1966}
David Hestenes.
\newblock {\em Space-time algebra}.
\newblock Birkhäuser, 1966.

\bibitem{Dressel2015}
J.~Dressel, K.~Y. Bliokh, and Franco Nori.
\newblock Spacetime algebra as a powerful tool for electromagnetism, 2015.

\bibitem{Doran2003}
C.~Doran and A.~Lasenby.
\newblock {\em Geometric Algebra for Physicists}.
\newblock Cambridge University Press, 2003.

\bibitem{Sobczyk2019}
Garret Sobczyk.
\newblock {\em Matrix Gateway to Geometric Algebra, Spacetime and Spinors}.
\newblock Independently published, 2019.

\bibitem{Gunn2016}
Charles Gunn.
\newblock Doing euclidean plane geometry using projective geometric algebra.
\newblock {\em Adv. Appl. Clifford Algebras}, 2016.

\bibitem{Sokolov2016}
Andrey Sokolov.
\newblock Clifford algebra and the projective model of hyperbolic spaces, 2016.

\bibitem{Roelfs2023}
M.~Roelfs and S.~De Keninck.
\newblock Graded symmetry groups: Plane and simple.
\newblock {\em Adv. Appl. Clifford Algebras}, 2023.

\bibitem{Hidden}
M.~Roelfs.
\newblock Hidden in the fold, 2023.

\bibitem{Hrdina2021}
J.~Hrdina and et~al.
\newblock Projective geometric algebra as a subalgebra of conformal geometric algebra.
\newblock {\em Adv. Appl. Clifford Algebras}, (31), 2021.

\bibitem{Neugebauer2018}
M.~Neugebauer and et~al.
\newblock Magnetic and electric transverse spin density of spatially confined light.
\newblock {\em Pysical Review X}, 2018.

\bibitem{Bliokh2014}
K.~Bliokh and et~al.
\newblock Extraordinary momentum and spin in evanescent waves.
\newblock {\em Nature communications}, 2014.

\bibitem{sobszyk2024}
G.~Sobszyk.
\newblock Geometric algebras of light cone projective graph geometries.
\newblock {\em Adv. Appl. Clifford Algebras}, (34), 2024.

\end{thebibliography}

%\printbibliography

%\xymatrix{
% & \eta \ar@{->}[ld]_{\mathrm{Lorentz}} \ar@/^/@{-^>}[rr]^{\mathrm{Raising}} \ar@{-->}[rdd]_{f_\pm} &  & x \ar@/^/@{-^>}[ll]^{\mathrm{Un-raising}} \ar@{->}[rd]^{\mathrm{Euclidean}} \ar@{-->}[ldd]^{f_\pm} &  \\
%\Lambda \ar@{-->}[rrd]_{f_\pm} &  &  &  & \Psi \ar@{-->}[lld]^{f_\pm} \\
% &  & \psi_{\mathrm{L}/\mathrm{R}}\in\mathbb{G}_3f_\pm &  & 
%}

% ------------------------------------------------------------------------
\end{document}